\documentclass{PoS}
\usepackage[english]{babel}
\usepackage[utf8]{inputenc}
\usepackage{cite}
\usepackage{xspace}
\usepackage[T1]{fontenc} 
\usepackage{subfig}
\usepackage{tabularx}  
\usepackage{longtable}  %
\usepackage{booktabs} %
\usepackage{units} %
\usepackage{comment}
\usepackage{pict2e}
\usepackage{lineno}

\def\LV{\ifmmode {\mathrm{LIV}}\else{\scshape LIV}\fi\xspace}
\def\LIV{\ifmmode{\mathrm{LIV}}\else{\scshape LIV}\fi\xspace}
\def\LI{\ifmmode {\mathrm{LI}}\else{\scshape LI}\fi\xspace}
\def\Ec{\ifmmode {\mathrm{E_c}}\else{\scshape $\rm E_c$}\fi\xspace}
\def\hEc{\ifmmode {\mathrm{\hat{E}_c}}\else{\scshape $\rm \hat{E}_c$}\fi\xspace}
\def\ELIV{\ifmmode {\mathrm{E_{\LIV}}}\else{\scshape $\rm E_{\LIV}$}\fi\xspace}
\def\ELIVl{\ifmmode {\mathrm{E_{\LIV}^{(1)}}}\else{\scshape $\rm E_{\LIV}^{(1)}$}\fi\xspace}

\title{Constraints on Lorentz invariance violation using HAWC observations above 100 TeV}

\ShortTitle{LIV constraints with HAWC}

\author{ 
    \speaker{H. Martínez-Huerta}$^{\rm a}$,
    S. Marinelli$^{\rm b}$, 
    J. T. Linnemann$^{\rm b}$
    and J. Lundeen$^{\rm b,c}$ 
    \ \ \ \ \ \ \ \ \ \ \ \ \ \ \ \ \ \ \ \ \ \ \ \ \ \ 
    for the HAWC Collaboration\footnote{
    For a complete author list, see http://www.hawc-observatory.org/collaboration/icrc2019.php 
    . For collaboration list see PoS(ICRC2019)1177.
    } \\ \\
    $^{\rm a}$ Instituto de Física de São Carlos, Universidade de São Paulo, São Carlos, SP, Brasil \\
    $^{\rm b}$ Department of Physics and Astronomy, Michigan State University, MI, USA \\ 
    $^{\rm c}$ Physics Division, Los Alamos National Laboratory, Los Alamos, NM, USA \\
        
        E-mail: \email{humbertomh@ifsc.usp.br}}

\abstract{Due to the high energies and long distances involved, astrophysical observations provide a unique opportunity to test possible signatures of Lorentz Invariance Violation (LIV). Superluminal LIV enables the decay of photons at high energy over relatively short distances, giving astrophysical spectra which have a hard cutoff above this energy. 
The High Altitude Water Cherenkov (HAWC) observatory is the most sensitive
currently-operating gamma-ray observatory in the world above 10 TeV.
Together with the recent development of an energy-reconstruction algorithm for HAWC using an artificial neural network, HAWC can make detailed measurements of gamma-ray ener\-gies above 100 TeV. With these observations, HAWC can limit the LIV energy scale greater than $10^{31}$ eV, over 800 times the Planck energy scale. 
This limit on LIV is over 60 times more constraining than the best previous value for $\ELIVl$.
}

\FullConference{36th International Cosmic Ray Conference -ICRC2019-\\
		July 24th - August 1st, 2019\\
		Madison, WI, U.S.A.}

\begin{document}


\section{Introduction}

The High Altitude Water Cherenkov  Observatory (HAWC) is a wide-field of view array of 300 tanks of 200,000 liters of water, each containing four photomultiplier tubes detectors. HAWC is located at 4100 m above sea level at 19º N near the Sierra Negra volcano, in Puebla, México and covers an area of 22,000 m$^2$. Since 2015, HAWC has operated over 95$\%$ duty cycle and is the most sensitive currently-operating gamma-ray observatory in the world above 10 TeV. 
In fact, HAWC recently reported detailed measurements of gamma-ray energies above 100 TeV \cite{HAWC_CRAB_2019} with the recent
development of advanced energy-reconstruction algorithms, including using an artificial neural network (see also Ref.~\cite{HAWC_ICRC19_Kelly} for the application to the Crab Nebula energy spectrum and Ref.~\cite{HAWC_ICRC19_Jim} in this Proceedings). 

Among the studies that this is generating, there is also the opportunity to test fundamental physics, such as the Lorentz invariance violation (\LIV), through the precise measurement and reconstruction of these unprecedented very-high-energy photons. 
Some effects of \LIV are expected to increase with energy and over very long distances due to cumulative processes in photon propagation. Therefore, astrophysical searches provide sensitive probes of LIV and its potential signatures, such as the energy-dependent time delay, photon splitting, vacuum Cherenkov radiation, photon decay, and many other phenomena~\cite{Martinez-Huerta:2016azo,Hohensee:2008xz,Coleman:1997xq, Klinkhamer:2008ky,Vasileiou:2013vra,Astapov:2019xmt}.

Previous studies of possible LIV constraints with HAWC have indicated its particle use in LIV searches. For instance, Ref. \cite{HAWC_LIV_GRB} analyzes the possibility to test energy-dependent time delays through GRB and Pulsar measurements, which would result in strong sensitivity limits to \LIV in the photon sector. In Ref. \cite{HAWC_LIV_PD}, the potential of Lorentz invariant violating photons to decay to electron-positron pairs was explored. Furthermore, preliminary results in this vein were presented in Ref. \cite{HAWC_LIV_CPT} and will be considered in the present proceeding. 

Superluminal LIV enables the decay of photons at high energy over relatively short distances and above the energy threshold of the process. Consequently, no high-energy photons should reach the Earth from astrophysical distances above some photon energy~\cite{Martinez-Huerta:2016azo}. Moreover, this suggests a hard cutoff in astrophysical spectra~\cite{Sam_Thesis}. In this work, seven sources are studied to determine whether or not there is a hard cutoff compatible with the LIV photon decay in the observed spectra of each source. In the next section, we present the highlights of the LIV photon decay phenomena. In Section 3, we describe the developed analysis and present our preliminary results, and finally, we present our conclusions.

\section{Lorentz invariance violation}

The introduction of a Lorentz violating term in the standard model Lagrangian or spontaneous Lorentz symmetry breaking can induce modifications to the particle dispersion relation, compared to the standard energy-momentum relationship in special relativity~\cite{Coleman:1997xq,Coleman:1998ti,Colladay:1998fq}.  Although there are several forms of modified dispersion relation (MDR) for different particles and underlying \LV-theories, some of them may lead to similar phenomenology, which can be useful for \LV tests in extreme environments such as astroparticle scenarios. Phenomenologically, the LIV effects can be generalized as a function of energy and momentum. In this way, a family of effective MDRs can be addressed by the following expression\footnote{Hereafter, natural units are used, $c=\hbar=1$.},
\begin{equation}\label{eq:GDR}
    E_{a}^{2} -  p_{a}^{2} = m_a^2 \pm |\alpha_{a,n}|A_a^{n+2},
\end{equation}
where $a$ stands for the particle type. $A$ can take the form of E or p. $\alpha_{a,n}$ is the \LV parameter and $n$ is the leading order of the correction from the underlying theory. In some effective field theories, $\alpha_{a,n}=\epsilon^{(n)}/M$, where $\epsilon^{(n)}$ are \LV coefficients and $M$ is the energy scale of the new physics, such as the Plank energy scale ($\rm E_{Pl}\approx 1.22 \times 10^{28}$) or some Quantum Gravity energy scale ($\rm E_{QG}$). 

The phenomenology derived from Eq.~(\ref{eq:GDR}) can demonstrate superluminal phenomena predicted in some \LIV scenarios, such as photon decay and vacuum Cherenkov radiation~\cite{Martinez-Huerta:2016azo}. In the photon decay scenario, the resulting decay rates into electron-positron are very fast and effective at energies where the process is allowed~\cite{Martinez-Huerta:2017ntv}. This creates a hard cutoff in the gamma-ray spectrum with no high-energy photons reaching the Earth from cosmological distances above a given threshold. The derived general threshold for any order~$n$ from Eq.~(\ref{eq:GDR}), is given by
\begin{equation}\label{eq:th}
    \alpha_{n} \ge \frac{4 m_e^2}{E_{\gamma}^{n}(E_{\gamma}^2-4 m_e^2)},
\end{equation}
where $E_{\gamma}$ is the photon energy and $m_e$ stands for the electron mass.

\section{Analysis and Lorentz violation limits}

\begin{figure}[t]
    \centering
    \subfloat[]{{\includegraphics[height=.32\linewidth]{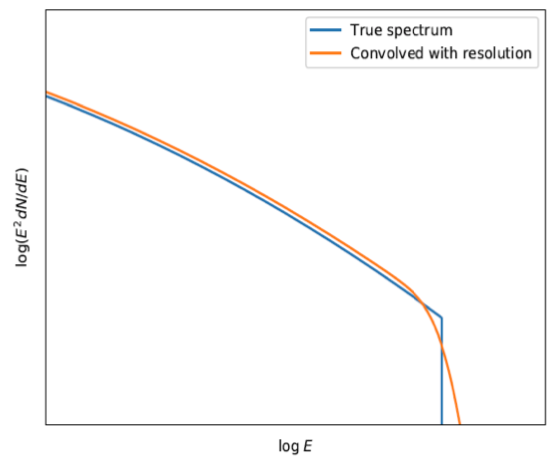} }}%
    \subfloat[]{{\includegraphics[width=.48\linewidth]{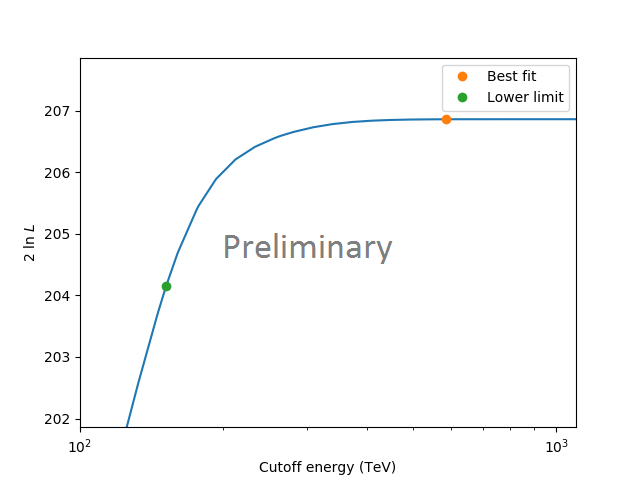} }}%
    \caption{\small Left (a). True spectrum with LIV hard cutoff at some energy $\rm E_c$ and the expected observer spectrum due to the detector energy resolution~\cite{Sam_Thesis}. Right (b). Likelihood curve as a function of the LIV Energy cutoff in the Crab analysis; the lower point (green) shows the lower limit at 95$\%$ CL.
    }
    \label{LIV:fig1}%
\end{figure}

To test LIV photon decay, the precise measurement of the most energetic gamma-ray events is crucial. HAWC is capable of detecting very high energy gamma-rays by studying astrophysical sources such as the Crab Nebula and other TeV sources. HAWC is also especially sensitive to sources with spatially-extended emission because of its large instantaneous field of view (2 sr). Furthermore, a recent Neural-Net-based (NN) energy estimation technique has improved the energy resolution of the instrument~\cite{HAWC_CRAB_2019}. 

A study of high-energy HAWC sources has been performed which uses a selection of seven sources with significant high energy emission above 56 TeV in reconstructed energy. They are the Crab Nebula,  modeled as a point source with a log-parabola spectrum, and six sources within 2 degrees of the Galactic plane, with spectra are shaped as a power-law with an exponential cut-off. These six sources are spatially extended and modeled by a Gaussian morphology with radius fixed at best-fit values found during the construction of HAWC's high-energy catalog~\cite{Abeysekara:2017hyn}. 

To test the LIV photon decay signature, we assume a LIV hard cutoff at some energy \Ec in the true spectrum, which is expected to be softened in the observed spectra due to the effects of the detector energy resolution (Fig.~\ref{LIV:fig1}~(a)).
A profile log-likelihood is performed to find the best-fit spectrum model for each source, including a energy cutoff, \hEc, as a free parameter. The LI scenario is recovered when $\hEc \rightarrow \infty$. The test-statistic between the best-fit LIV and LI is given by  
\begin{equation}
D = 2\ln \left(\frac{\mathcal{L}(\hEc)}{\mathcal{L}(\hEc \rightarrow \infty)}\right).
\end{equation}

To determine whether or not the data is compatible with a hard cutoff at some energy scale, we calculate a p-value of observing $D$, where the null hypothesis is the LI limit.  The resulting p-value for each of the seven sources is shown in Tab.~\ref{tab:1}.  As can be seen, the p-values are large enough that the null hypothesis cannot be rejected; therefore, none of the sources favor a spectrum with an LIV hard cutoff. Instead, we consider limits to how high in energy the presence of photons can be verified with HAWC.
In this analysis, it is found that the lower limit to \Ec at 95$\%$ CL occurs where $2\ln \mathcal{L}(\Ec)$ changes by $2.71$ units from the best fit. An example is given in Fig.~\ref{LIV:fig1}~(b), which shows the likelihood curve as a function of the energy cutoff for the Crab source. The top point (orange) shows the best-fit value and the lower point (green) is the lower limit at 95$\%$ CL. The lower limit results for $\rm E_c$  at 95$\%$ for all sources are presented in Tab 1. Then, by using Eq.~(\ref{eq:th}), the 95$\%$ CL the limits are reinterpreted as limits on LIV parameters in Tab.~\ref{tab:1}. In this way, HAWC can exclude the energy scale of the new LIV physics, $\ELIVl$, to greater than $10^{31}$eV, over 800 times the Planck energy scale and 60 times more constraining than the best previous value.
 Previous strong limits testing photon decay using very-high energy photons from HEGRA telescope \cite{Martinez-Huerta:2016azo}, Tevatron \cite{Hohensee:2008xz}, and HESS \cite{Klinkhamer:2008ky} are given for comparison in Tab.~\ref{tab:2} \footnote{where $\alpha_{n} =\rm E_{\LV}^{(-n)} \approx \rm E_{\rm QG}^{(-n)}$, and in the framework of the Standard Model Extension~\cite{Colladay:1998fq},  $\alpha_0 \approx  -2 \widetilde{\kappa}_{\rm tr}$ and $\alpha_{2} = - c_{(I)0,0}^{(6)} / \sqrt{\pi}$ .}. Limits due to LIV energy-dependent time delay searches with the {\it Fermi}-LAT are also shown~\cite{Vasileiou:2013vra}, as well as the limits due to  superluminal photon splitting~\cite{Astapov:2019xmt}. 

\begin{table}[th!]
\small \centering
\begin{tabular}{@{}lcccccll@{}} \toprule
\multicolumn{1}{c}{Source} & \begin{tabular}[c]{@{}c@{}}$\rm E_c $\\  TeV \end{tabular} & \begin{tabular}[c]{@{}c@{}}$|\alpha_0|$\\  $10^{-17}$\end{tabular} & \begin{tabular}[c]{@{}c@{}}$|\alpha_1|$ \\ $10^{-31}$eV$^{-1}$\end{tabular} & \begin{tabular}[c]{@{}c@{}}$|\alpha_2|$\\ $10^{-45}$eV$^{-2}$\end{tabular} & \begin{tabular}[c]{@{}c@{}}$\rm E_{\LV}^{(1)}$\\  $10^{30}$eV\end{tabular} & \begin{tabular}[c]{@{}c@{}}$\rm E_{\LV}^{(2)}$\\ $10^{22}$eV\end{tabular} &
\begin{tabular}[c]{@{}c@{}} p\\ value \end{tabular}\\ \toprule
2HWC J1825-134 \ \ \ \ \ \ \ \ \ \ \ \ \ \ \ \ \ \ \ \ \ \ \ & 253 & 1.63 & 0.64 & 0.26 & 15.5 & 6.26 & 1\\
2HWC J1908+063  &   213 &   2.30    &   1.08    &   0.51    &   9.25    & 4.44  &   0.99 \\ 
Crab (HAWC) & 152 & 4.52 &	2.97 & 1.96 & 3.4 & 2.26 & 1 \\ 
2HWC J2031+415 & 144 & 5.04 & 3.5 & 2.43 & 2.9 & 2.02 & 0.714 \\
2HWC J2019+367 & 121 & 7.13 & 5.6 & 4.87 & 1.7 & 1.43 & 0.828  \\
J1839-057 & 79 & 16.74 & 21.1 & 26.8 & 0.47 & 0.61 & 0.357 \\ 
2HWC J1844-032 & 77 &17.62 & 22.9 & 29.7 & 0.44 & 0.58 & 0.294  \\
\bottomrule
\end{tabular}
\caption{The HAWC Sources used in this analysis and the derived 95$\%$ CL lower limits on $\rm E_c$ and its different LIV coefficients (Prel.).
}\label{tab:1}

\end{table}

\begin{table}[th!]
\small \centering
\begin{tabular}{@{}lcccccll@{}} \toprule
\multicolumn{1}{c}{Source} & \begin{tabular}[c]{@{}c@{}}$\rm E_\gamma $\\  TeV \end{tabular} & \begin{tabular}[c]{@{}c@{}}$|\alpha_0|$\\  $10^{-17}$\end{tabular} & \begin{tabular}[c]{@{}c@{}}$|\alpha_1|$ \\ $10^{-31}$eV$^{-1}$\end{tabular} & \begin{tabular}[c]{@{}c@{}}$|\alpha_2|$\\ $10^{-45}$eV$^{-2}$\end{tabular} & \begin{tabular}[c]{@{}c@{}}$\rm E_{\LV}^{(1)}$\\  $10^{30}$eV\end{tabular} & \begin{tabular}[c]{@{}c@{}}$\rm E_{\LV}^{(2)}$\\ $10^{22}$eV\end{tabular} &
\begin{tabular}[c]{@{}c@{}} Ref.\\ \end{tabular}\\ \toprule
Crab (HEGRA) 2017 & $\sim 56$ & - & 66.7 & 128 & 0.15 & 0.28 & \cite{Martinez-Huerta:2016azo} \\
Tevatron 2016
& 0.442 & $6\times10^{5}$ & - & - & - & - & \cite{Hohensee:2008xz} \\
RX J1713.7–3946 (HESS) 2008
& 30 & 180 & - & - & - & - & \cite{Klinkhamer:2008ky} \\
Coleman $\&$ Glashow (1997) 
& 20 & 100 & - & - & - & - & \cite{Coleman:1997xq}   \\
\midrule
GRB09510 ({\it Fermi}) 2013 $v>c$ & - & - & - & - & 0.134 & 0.009 & \cite{Vasileiou:2013vra}\\
GRB09510 ({\it Fermi}) 2013 $v<c$ & - & - & - & - & 0.093 & 0.013 & \cite{Vasileiou:2013vra}\\
\midrule
Crab (HEGRA) 2019 & 75 & - & - & 0.059 & - & 13  & \cite{Astapov:2019xmt}\\
\bottomrule
\end{tabular}
\caption{Previous strong constraints to LIV photon decay are shown as well as the best limits based on energy-dependent time delay and superluminal photon splitting at bottom.
}\label{tab:2}
\end{table}

\section{Conclusions}

The HAWC observatory measurements of the highest-energy photons can be used as a test to probe fundamental physics such as Lorentz violation. In this work, we set preliminary \LIV limits by testing the LIV photon decay through the study of seven sources with significant high energy emission, including the Crab Nebula. It was found that none of them favor a spectrum with a hard cutoff. However, the dedicated search of such signature in the spectra increases the energy to which the existence of the most energetic photons can be confirmed, which leads to new and stringent limits to \LIV. A study including detailed systematic uncertainties in the source spectra and HAWC detector response will be addressed in a future publication. 


\section*{Acknowledgments}
\footnotesize{ 
We acknowledge the support from: the US National Science Foundation (NSF); the US~Department of Energy Office of High-Energy Physics; the Laboratory Directed Research and De\-ve\-lopment (LDRD) program of Los Alamos National Laboratory; Consejo Nacional de Ciencia y Tecnolog\'{\i}a (CONACyT), M{\'e}xico (grants 271051, 232656, 260378, 179588, 239762, 254964, 271737, 258865, 243290, 132197), Laboratorio Nacional HAWC de rayos gamma; L'OREAL Fellowship for Women in Science 2014; Red HAWC, M{\'e}xico; DGAPA-UNAM (grants IG100317, IN111315, IN111716-3, IA102715, 109916, IA102917); VIEP-BUAP; PIFI 2012, 2013, PROFOCIE 2014, 2015;the University of Wisconsin Alumni Research Foundation; the Institute of Geophysics, Planetary Physics, and Signatures at Los Alamos National Laboratory; Polish Science Centre grant DEC-2014/13/B/ST9/945; Coordinaci{\'o}n de la Investigaci{\'o}n Cient\'{\i}fica de la Universidad Michoacana.  Thanks to Luciano D\'{\i}az and Eduardo Murrieta for technical support. 
HMH acknowledges FAPESP support No. 2015/15897-1 and 2017/03680-3 and the National Laboratory for Scientific Computing (LNCC/MCTI, Brazil) for providing HPC resources of the SDumont supercomputer (\href{https://sdumont.lncc.br}{sdumont.lncc.br}).
}

    

\end{document}